\documentclass[11pt,epsf]{article}
\usepackage{amsmath}
\usepackage{amsfonts}
\usepackage{amssymb}
\usepackage{graphicx}

\newcommand {\dfn} {\stackrel{\Delta} {=}}

\newcommand {\eqas} {\stackrel{\mbox{a.s.}} {=}}

\newcommand {\bp} {\mbox{\boldmath $p$}}

\newcommand {\br} {\mbox{\boldmath $r$}}

\newcommand {\bw} {\mbox{\boldmath $w$}}
\newcommand {\bx} {\mbox{\boldmath $x$}}

\newcommand {\bE} {\mbox{\boldmath $E$}}

\newcommand {\bX} {\mbox{\boldmath $X$}}

\newcommand{\calC}{{\cal C}}

\newcommand{\calE}{{\cal E}}

\newcommand{\calX}{{\cal X}}
\newcommand{\calY}{{\cal Y}}

\begin{document}
\thispagestyle{empty}
\title{On the Statistical Physics of Directed Polymers in a Random Medium
and Their Relation to Tree Codes\thanks{
This research was supported by the Israel Science Foundation (ISF),
grant no.\ 208/08.}}
\author{Neri Merhav}
\date{}
\maketitle

\begin{center}
Department of Electrical Engineering \\
Technion - Israel Institute of Technology \\
Haifa 32000, ISRAEL \\
{\tt merhav@ee.technion.ac.il}\\
\end{center}
\vspace{1.5\baselineskip}
\setlength{\baselineskip}{1.5\baselineskip}

\begin{abstract}
Using well--known results from statistical physics, 
concerning the almost--sure behavior of the free energy of
directed polymers in a random medium, we prove that random tree codes
achieve the distortion--rate function almost surely under a
certain symmetry condition.

\vspace{0.5cm}

\noindent
{\bf Index Terms:} directed polymer, Cayley tree, 
free energy, partition function, 
tree coding, rate--distortion theory, delay.
\end{abstract}

\newpage
\section{Introduction}

Tree source coding with a fidelity criterion has been studied since the late
sixties and the early seventies of the previous century,
see, e.g., \cite[Subsection
6.2.4]{Berger71},\cite{DH75},\cite{DBJ74},\cite{Gallager74},\cite{Jelinek69},\cite{JA71}.
The first results, that were obtained by Jelinek and Anderson \cite{JA71}, were
for tree coding of binary sources with the Hamming distortion measure, and by
Dick, Berger and Jelinek \cite{DBJ74} for
Gaussian sources and the squared error distortion measure. Davis and Hellman
\cite{DH75} proved a tree coding theorem for a general memoryless 
source and a general fidelity
criterion. In particular, they pointed out that in an earlier paper by Jelinek
\cite{Jelinek69}, the proof of the coding theorem was valid only for symmetric
sources, and so, by modifying the 
branching process associated with the tree code,
they were able to relax the symmetry 
condition of the tree coding theorem. In this context, it
should be pointed out that Gallager \cite{Gallager74} 
also made a symmetry assumption in the same spirit.

The main message in this short paper is, first of all, in the observation
that the tree source coding problem is very intimately related to an important
model in statistical physics of disordered systems, namely,
the directed polymer in a random medium (DPRM), cf.\ e.g.,
\cite{Bhattacharjee08},\cite{BPP93},\cite{CSY04},\cite{CD91},\cite{Derrida90},\cite{DES93},\cite{ED92},\cite{MG07},\cite{PdLS98}
and references therein. 
Loosely speaking, in the DPRM, each 
configuration of the underlying physical system
corresponds to a walk along consecutive bonds of a certain lattice, or a tree,
where each such bond is assigned with an 
independent random variable (energy), and
where the
total energy (which is analogous to the distortion of the tree code)
of this walk is the sum of energies along the bonds visited. For
a given realization of these random energy variables, the probability
of each walk is given by the Boltzmann distribution, namely, it is
proportional
to an exponential function of the negative total energy. The main challenge,
as usual in equilibrium statistical physics, is to characterize the asymptotic
normalized free energy
of a typical realization of the system. For the case where the walks are
defined on a tree (from the root to one of the leaves), this problem has
a closed--form solution.

This relationship between tree codes and the DPRM 
is interesting on its own right. It turns out to be so strong,
that the various 
analysis techinques\footnote{These techniques are different from those of 
the papers mentioned in the first paragraph.}
and the results concerning the DPRM can readily be harnessed 
to the ensemble peformance analysis of tree codes. In particular,
the distortion achieved by the best codeword in the tree codebook is
identified with the free energy of 
the DPRM when the system is frozen (taken to zero
temperature). This observation, 
does not merely provide an alternative proof of the
tree coding theorem, but moreover, it enables to show that,
at least under a certain symmetry assumption
concerning the source and the distortion function\footnote{This assumption is
in the spirit of the above mentioned assumption by Gallager, though it is
somewhat different.}, the distortion--rate function is achieved eventually
almost surely (with respect to the randomness of the code) for every
individual source
sequence. This is different from (and stronger than) 
the previous findings, mentioned in the first
paragraph above, which were coding theorems concerning the average distortion.

The outline of this work is as follows: In Section 2, we establish our notation
conventions and give a brief background in statistical mechanics
in general and on the DPRM in particular. In Section 3, we show how 
the solution to the DPRM model can be used to prove that the tree
code ensemble achieves distortion--rate function almost surely for every input.
Finally, in Section 4, we provide a short summary of this paper.

\section{Notation Conventions and Background}
\subsection{Notation Conventions}

Throughout this paper, scalar random
variables (RV's) will be denoted by capital
letters, like $X$ and $Y$, their sample values will be denoted by
the respective lower case letters, and their alphabets will be denoted
by the respective calligraphic letters.
A similar convention will apply to
random vectors and their sample values,
which will be denoted with the same symbols in the boldface font.
Thus, for example, $\bX$ will denote a random $n$-vector $(X_1,\ldots,X_n)$,
and $\bx=(x_1,...,x_n)$ is a specific vector value in $\calX^n$,
the $n$-th Cartesian power of $\calX$.
Sources and other probability measures
that underly sequence generation will be denoted generically by the letters $P$ and $Q$,
and specific letter probabilities will be denoted by the corresponding lower
case letters, e.g., $p(x)$, $q(y)$, etc.
The expectation operator will be
denoted by $\bE\{\cdot\}$. 
Information theoretic quantities like entropies and mutual
informations will be denoted following the usual conventions
of the Information Theory literature.

\subsection{Background}
\label{phys}

Consider a physical system with $n$ particles,
which can be in a variety of microscopic states (`microstates'),
defined by combinations of physical quantities associated with
these particles, e.g.,
positions, momenta,
angular momenta, spins, etc., of all $n$ particles.
For each such
microstate of the system, which we shall
designate by a vector $\bx=(x_1,\ldots,x_n)$, there is an
associated energy, given by an {\it Hamiltonian} (energy function),
$\calE(\bx)$. For example, if $x_i=(\bp_i,\br_i)$, where
$\bp_i$ is the momentum vector of particle number $i$ and
$\br_i$ is its position vector, then classically, $\calE(\bx)=\sum_{i=1}^N
[\frac{\|\bp_i\|^2}{2m}+mgz_i]$, where $m$ is the mass of each particle,
$z_i$ is its height -- one of the
coordinates of $\br_i$, and
$g$ is the gravitation constant.

One of the most
fundamental results in statistical physics (based on the
law of energy conservation and
the basic postulate that all microstates of
the same energy level are equiprobable)
is that when the system is in thermal
equilibrium with its environment, the probability of finding the system in
a microstate $\bx$ is
given by the {\it Boltzmann--Gibbs} distribution
\begin{equation}
\label{bd}
P(\bx)=\frac{e^{-\beta\calE(\bx)}}{Z(\beta)}
\end{equation}
where $\beta=1/(kT)$, $k$ being Boltzmann's contant and $T$ being temperature,
and
$Z(\beta)$ is the normalization constant,
called the {\it partition function}, which
is given by
$$Z(\beta)=\sum_{\bx} e^{-\beta\calE(\bx)}$$
or
$$Z(\beta)=\int \mbox{d}\bx e^{-\beta\calE(\bx)},$$
depending on whether $\bx$ is discrete or continuous. The role
of the partition function is by far deeper
than just being a normalization factor, as
it is actually the key quantity from which many
macroscopic physical quantities can be derived,
for example, the free energy\footnote{The free
energy means the maximum work that the system
can carry out in any process of fixed temperature.
The maximum is obtained when the process is reversible (slow, quasi--static
changes in the system).}
is $F(\beta)=-\frac{1}{\beta}\ln Z(\beta)$, the average internal
energy (i.e., the expectation of $\calE(\bx)$ where
$\bx$ drawn is according (\ref{bd}))
is given by $\bar{E}\dfn\bE\{\calE(\bX)\}=-(\mbox{d}/\mbox{d}\beta)\ln
Z(\beta)$, the heat capacity
is obtained from
the second derivative, etc. One of the ways to
obtain eq.\ (\ref{bd}), is as the maximum entropy
distribution under an average energy constraint
(owing to the second law of thermodynamics), where $\beta$ plays the role of
a Lagrange multiplier that controls the average energy.

Quite often, real--world physical systems of many particles, 
such as magnetic materials and solid--state devices, are subjected to 
effects of impurity (e.g., defects) that 
may appear as amorphic structures and disorder.
To model such disorder, it is customary to let the Hamiltonian,
$\calE(\bx)$, depend also on certain random parameters and to examine the
behavior of systems pertaining to typical realizations of these random
parameters. There are many models
of this kind in the physics literature. One of them is the DPRM, which is
defined on a certain graph, such as a hypercubic lattice, or a tree. We
henceforth focus on the latter and describe it more formally than in the
Introduction.

\begin{figure}[ht]
\hspace*{5cm}\input{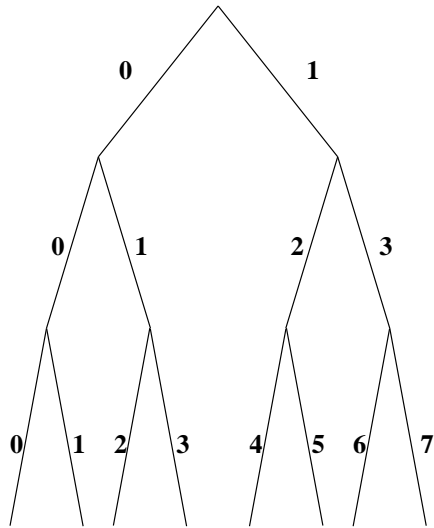}
\caption{A Cayley tree with branching factor $d=2$ and depth $n=3$.}
\label{tree}
\end{figure}

Consider a {\it Cayley tree}, namely, a full balanced tree with branching ratio
$d$ and depth $n$ (cf.\ Fig.\ \ref{tree}, where $d=2$ and $n=3$). Let us index the
branches by a pair of integers $(i,j)$, where $1\le i\le n$ describes the
generation (with $i=1$ corresponding to the $d$ branches that emanate from the root),
and $0\le j\le d^i-1$ enumerates the branches of the $i$--th generation, say,
from left to right (see Fig.\ \ref{tree}). 
For each branch $(i,j)$, $1\le j\le d^i$, $1\le i\le n$,
we randomly draw an independent random variable $\varepsilon_{i,j}$ according
to a fixed probability function $q(\varepsilon)$ (i.e., a probability mass function in the discrete
case, or probability density function in the continuous case).

A {\it walk} $\bw$, from the root of the tree to one of its leaves, is described
by a finite
sequence $\{(i,j_i)\}_{i=1}^n$, where $0\le j_1\le d-1$ and $dj_i\le
j_{i+1}\le dj_i+d-1$, $i=1,2,\ldots,(n-1)$.\footnote{In fact, for a given $n$, the
number $j_n$ alone dictates the entire walk.} For a given realization of the RV's
$\{\varepsilon_{i,j}:~i=1,2,\ldots,n,~j=0,1,\ldots,d^i-1\}$, we define the
Hamiltonian associated with $\bw$ as
$\calE(\bw)=\sum_{i=1}^n\varepsilon_{i,j_i}$, 
and then the partition function as:
\begin{equation}
\label{pftree}
Z_n(\beta)=\sum_{\bw} \exp\{-\beta\calE(\bw)\}.
\end{equation}
Of course, since $\{\varepsilon_{i,j}\}$ are RV's, then so is $Z_n(\beta)$.
The primary question addressed by physicists, in this context,
concerns the (typical) behavior of the RV 
\begin{equation}
f_n(\beta)\dfn\frac{1}{n\beta}\ln Z_n(\beta) 
\end{equation}
for $n$ large,
which is (up to the minus sign), exactly the normalized free energy per step. It
turns out (as proved e.g., in \cite{BPP93},\cite{DES93}) that $f_n(\beta)$
has a {\it self--averaging} property, in the terminology of physicists, in
other words, the sequence of random variables $\{f_n(\beta)\}_{n\ge 1}$ 
converges in probability (and in fact, almost surely, as is shown in
\cite{BPP93}) to a deterministic
constant $f(\beta)$, which is given by
\begin{equation}
\label{physresult}
f(\beta)=\left\{\begin{array}{ll}
\phi(\beta) & \beta\le\beta_c\\
\phi(\beta_c) & \beta>\beta_c
\end{array}\right.
\end{equation}
with the function $\phi$ being defined as
\begin{equation}
\phi(\beta)\dfn \frac{\ln[d\cdot\bE\{e^{-\beta \varepsilon}\}]}{\beta}
\end{equation}
where the expectation, which is assumed finite, is taken 
w.r.t.\ $q(\varepsilon)$, and where $\beta_c$ is
the value of $\beta$ at which $\phi(\beta)$ is minimum, or equivalently,
the solution to the equation $\phi'(\beta)=0$, where $\phi'$ is the derivative
of $\phi$. 

As can be seen, $\beta=\beta_c$ is a point at which the asymptotic
normalized free energy per step, $f(\beta)$, 
changes its behavior: Although $f(\beta)$ and
its first derivative are continuous functions for all $\beta$, the second
derivative is discontinuous at $\beta=\beta_c$. In the terminology of
physicists, this is referred to as a {\it second order phase transition}.
Observe that while one might expect that the sequence 
$f_n(\beta)$ would converge to
the same limit as $\bE\{f_n(\beta)\}=\frac{1}{n\beta}\bE\{\ln Z_n(\beta)\}$,
i.e., the so called {\it quenched average},
the high temperature phase result (\ref{physresult}) corresponds to
$\frac{1}{n\beta}\ln[\bE\{Z_n(\beta)\}]$, which is called the {\it annealed
average}. This means
that Jensen's inequality
is essentially tight at this range of $\beta$. However, these two averages
depart from each other at the low temperature phase, $\beta > \beta_c$. As can
be observed, in this phase, the asymptotic normalized 
free energy no longer depends on $\beta$,
and it is referred to as the {\it glassy phase} or the {\it frozen phase},
which is characterized by zero thermodynamical entropy, in other words, the partition
function is dominated by a sub--exponential number of configurations possessing
the ground--state energy (cf.\ e.g., \cite[Chap.\ 5]{MM06}). 
For reasons that will become apparent shortly, this frozen phase
is the relevant phase for our source coding problem.

The asymptotic free energy formula (\ref{physresult}) 
has been proved in the physics literature at least in four different ways:
The first \cite{BPP93} is based on martingales, the second is based on
non--integer moments of the partition function \cite{DES93},\cite{ED92},
the third is based on a recursion of a certain generating function 
of the partition function as well as on traveling waves \cite{Derrida90},\cite{DS88},
and the fourth method is the so--called {\it replica method} \cite{Derrida90},
which, although not rigorous, is very useful in statistical mechanics.

\section{Main Result}

We now turn to our lossy source coding problem, where some of the notation
that will be used will be deliberately identical to that of 
Subsection \ref{phys}. Consider a discrete memoryless
source (DMS) $P$ that generates symbols $X_1,X_2,\ldots$ from 
a finite\footnote{
Finite alphabet assumptions are made mostly for simplicity. It is expected
that our derivations continue to hold in the continuous case as well under
suitable regularity conditions.} alphabet
$\calX$. Let $\calY$ denote a finite reproduction alphabet and let
$\rho:\calX\times\calY\to[0,\infty)$ be a given distortion function.

Consider next an ensemble 
of tree codes for encoding source $n$--tuples,
$\bx=(x_1,\ldots,x_n)$, which is 
defined as follows: Given a coding rate $R$ (in nats/source--symbol), which is
assumed to be the natural logarithm of some positive integer\footnote{At first
sight, it might appear that this gives a rather limited variety of coding
rates to work with. Obviously, this can be improved by working with a
superalphabet of (small) blocks, as was done in previous works on tree coding.
But since the source alphabet could have been defined for these blocks in the first place,
there is no essential loss of generality in this setting.}
$d$, and given a
probability distribution on the reproduction alphabet, $Q=\{q(y),~y\in\calY\}$, 
let us draw $d=e^R$ independent copies of $Y$ under $Q$, and denote them by
$Y_1,Y_2,\ldots,Y_d$. We shall refer to the randomly chosen set,
$\calC_1=\{Y_{1},Y_{2},\ldots,Y_{d}\}$, as our `codebook' for the first
source symbol, $X_1$. Next, for each $1\le j_1\le d$, we randomly select another
such codebook under $Q$, 
$\calC_{2,j_1}=\{Y_{j_1,1},Y_{j_1,2},\ldots,Y_{j_1,d}\}$, for the second symbol,
$X_2$. Then, for each $1\le j_1\le d$ and $1\le j_2\le d$, we again draw under $Q$
yet another codebook
$\calC_{3,j_1,j_2}=\{Y_{j_1,j_2,1},Y_{j_1,j_2,2},\ldots,Y_{j_1,j_2,d}\}$, for $X_3$, and so on.
In general, for each $t\le n$, we randomly draw $d^{t-1}$ codebooks under
$Q$, which are indexed by $(j_1,j_2,\ldots,j_{t-1})$, $1\le j_k\le d$, $1\le k\le t-1$.

Once the above described random code selection process is complete, the resulting 
set of codebooks $\{\calC_1,\calC_{t,j_1,\ldots,j_{t-1}},~2\le t\le n,~1\le j_k\le
d,~1\le k\le t-1\}$
is revealed to both the
encoder and decoder, and the encoding--decoding system works as follows:
\begin{itemize}
\item {\it Encoding:} Given a source $n$--tuple $X^n$, find a vector of
indices $(j_1^*,j_2^*,\ldots,j_n^*)$ that minimizes the overall distortion
$\sum_{t=1}^n \rho(X_t,Y_{j_1,\ldots,j_t})$. Represent each component $j_t^*$
(based on $j_{t-1}^*$)
by $R=\ln d$ nats (that is, $\log_2d$ bits), thus a total of $nR$ nats.
\item {\it Decoding:}
At each time $t$ ($1\le t\le n$), after having decoded
$(j_1^*,\ldots,j_t^*)$, output the reproduction symbol
$Y_{j_1^*,\ldots,j_t^*}$.
\end{itemize} 
A few comments are in order at this point: First, as we see, the codebook generation process
is branching hierarchically by a factor of $d$ at each step, hence it is convenient to think
of the code as having the structure of
a Cayley tree, as in Subsection \ref{phys}. 
The encoder seeks the best walk on that tree
in the sense of minimum distortion. Note also that the process of converting
the optimum walk $\bw^*=(j_1^*,j_2^*,\ldots,j_n^*)$ into a compressed bitstream is
extremely simple: We just convert each $j_t\in\{1,\ldots,d\}$ into its binary
representation using $\log_2d$ bits without any attempt at compression. In
other words, the entropy coding part is trivial in the sense that it uses neither
the memory that may be present in the sequence $(j_1^*,j_2^*,\ldots,j_n^*)$,
nor the possible skewdness of the distributions of these symbols. Finally, the
decoding process is a purely sequential delayless process: 
At time $t$, the decoder outputs the $t$-th reproduction symbol. This is in
contrast to the decoder
of a general block code, which has to wait until the entire bit
string of length $nR$ has been received before it can start to decode. Thus, at
least the decoding delay is saved this way. There is also a slight reduction
in the search complexity at the encoder, due to the tree structure, but not a
dramatic one.

In order to analyze the rate--distortion performance of this ensemble of
codes, using the results of Subsection \ref{phys},
we now make the following assumption:

\vspace{0.15cm}

\noindent
{\it The random coding distribution $Q$ is such that the
distribtion of the RV $\rho(x,Y)$ is the same for all $x\in\calX$.}

It turns out that this assumption is fulfilled 
quite often -- it is the case whenever the random
coding distribution together with distortion function exhibit a sufficiently
high degree of symmetry. For example, if $Q$ is the uniform distribution over
$\calY$ and the rows of the distortion matrix $\{\rho(x,y)\}$ are permutations
of each other, which is in turn the case, for example, 
when $\calX=\calY$ is a group and $\rho(x,y)=\gamma(x-y)$
is a difference distortion function w.r.t.\ the group difference operation. 
Somewhat more generally, this assumption
still holds when the different rows of the
distortion matrix
are formed by permutations of each other subject to the following
rule: $\rho(x,y)$ can be
swapped with $\rho(x,y')$ provided that $q(y')=q(y)$.

It should be pointed out that if the optimum random coding distribution $Q^*$,
namely, the one corresponding to the output of the test channel that achieves
the rate--distortion function of $X$, happens to satisfy the above symmetry
assumption, then as we show below (using a technique different from
those of the earlier papers on tree coding),
the rate--distortion performance of the
above descrirbed code ensemble achieves the rate--distortion function.
Moreover, this will turn out to be the case, not only in expectation, but also
with probability one. 

We now turn to our analysis which makes heavy use of the results of 
Subsection \ref{phys}. 
For a given $\bx$ and a given realization of the set of codebooks, define
the partition function in analogy to that of the DPRM:
\begin{equation}
Z_n(\beta)=\sum_{\bw} \exp\{-\beta\sum_{t=1}^n\rho(x_t,Y_{j_1,\ldots,j_t})\},
\end{equation}
where the summation extends over all $d^n$ possible walks,
$\bw=(j_1,\ldots,j_n)$,
along the Cayley tree, as defined in Subsection \ref{phys}.
Clearly, considering our symmetry assumption, 
this falls exactly under the umbrella of the DPRM, with the distortions
$\{\rho(x_t,Y_{j_1,\ldots,j_t})\}$ playing the role of the branch energies
$\{\varepsilon_{i.j}\}$. Therefore, $\frac{1}{n\beta}\ln Z_n(\beta)$ converges
almost surely, as $n$ grows without bound, to $f(\beta)$, now defined as
\begin{equation}
\label{physresult1}
f(\beta)=\left\{\begin{array}{ll}
\phi(\beta) & \beta\le\beta_c\\
\phi(\beta_c) & \beta>\beta_c
\end{array}\right.
\end{equation}
where
\begin{eqnarray}
\phi(\beta)&\dfn&\frac{\ln[d\cdot\bE\{e^{-\beta\rho(x,Y)}\}]}{\beta}\nonumber\\
&=&\frac{\ln[e^R\cdot\bE\{e^{-\beta\rho(x,Y)}\}]}{\beta}\nonumber\\
&=&\frac{R+\ln[\bE\{e^{-\beta\rho(x,Y)}\}]}{\beta},
\end{eqnarray}
where $x$ is an arbitrary member of $\calX$, which is immaterial by the
symmetry assumption.
Thus, for every $(x_1,x_2,\ldots)$, the distortion is
given by
\begin{eqnarray}
\limsup_{n\to\infty}\frac{1}{n}\sum_{t=1}^n\rho(x_t,Y_{j_1^*,\ldots,j_t^*})
&\dfn&\limsup_{n\to\infty}\frac{1}{n}\min_{\bw}
\left[\sum_{t=1}^n\rho(x_t,Y_{j_1,\ldots,j_t})\right]\nonumber\\
&=&\limsup_{n\to\infty}\limsup_{\ell\to\infty}\left[-\frac{\ln
Z_n(\beta_\ell)}{n\beta_\ell}\right]\nonumber\\
&\le&\limsup_{\ell\to\infty}\limsup_{n\to\infty}\left[-\frac{\ln
Z_n(\beta_\ell)}{n\beta_\ell}\right]\nonumber\\
&\eqas&-\liminf_{\ell\to\infty}f(\beta_\ell)\\
&=& -\phi(\beta_c)\nonumber\\
&=&\max_{\beta\ge
0}\left[-\frac{\ln[\bE\{e^{-\beta\rho(x,Y)}\}]+R}{\beta}\right]\nonumber\\
&\dfn& D_0(R),
\end{eqnarray}
where: (i) $\{\beta_\ell\}_{\ell\ge 1}$ is an arbitrary sequence tending to
infinity, (ii) the almost--sure equality is due to \cite[Theorem
1]{BPP93}, and (iii) 
the inequality at the third line is
justified by the following chain:
\begin{eqnarray}
\limsup_{n\to\infty}\limsup_{\ell\to\infty}\left[-\frac{\ln
Z_n(\beta_\ell)}{n\beta_\ell}\right]&\le&
\limsup_{n\to\infty}\limsup_{\ell\to\infty}\left[-\frac{\ln\exp\{-\beta_\ell
\sum_{t=1}^n\rho(x_t,Y_{j_1^*,\ldots,j_t^*})\}}{\beta_\ell n}\right]\nonumber\\
&=&
\limsup_{n\to\infty}\frac{1}{n}
\sum_{t=1}^n\rho(x_t,Y_{j_1^*,\ldots,j_t^*})\nonumber\\
&=&\limsup_{\ell\to\infty}
\limsup_{n\to\infty}\frac{1}{n}
\sum_{t=1}^n\rho(x_t,Y_{j_1^*,\ldots,j_t^*})\nonumber\\
&=&\limsup_{\ell\to\infty}
\limsup_{n\to\infty}\left[-\frac{\ln\exp\{-\beta_\ell
\sum_{t=1}^n\rho(x_t,Y_{j_1^*,\ldots,j_t^*})\}}{\beta_\ell n}\right]\nonumber\\
&\le&\limsup_{\ell\to\infty}
\limsup_{n\to\infty}\left[-\frac{\ln[d^{-n}\sum_{\bw}\exp\{-\beta_\ell
\sum_{t=1}^n\rho(x_t,Y_{j_1,\ldots,j_t})\}}{\beta_\ell n}\right]\nonumber\\
&=&\limsup_{\ell\to\infty}\left\{
\limsup_{n\to\infty}\left[-\frac{\ln Z_n(\beta_\ell)}
{\beta_\ell n}\right]+\frac{\ln d}{\beta_\ell}\right\}\nonumber\\
&=&\limsup_{\ell\to\infty}
\limsup_{n\to\infty}\left[-\frac{\ln Z_n(\beta_\ell)}
{\beta_\ell n}\right]
\end{eqnarray}
We have shown then that the almost--sure distortion performance is uniformly given by
$D_0(R)$ for every individual
source sequence $x_1,x_2,\ldots$. Now, let us suppose that $Q$ is chosen to be the output distribution
$Q^*$ induced by the source $P$ and the test channel $\calX\to\calY$ that achieves
the rate--distortion function, and that the symmetry assumption continues to
hold for $Q^*=\{q^*(y),~y\in\calY\}$. Then, we claim that $D_0(R)$, defined
with $Q=Q^*$, coincides with the
distortion--rate function of the source, $D(R)$. 

To see why this is true, recall that the rate--distortion function $R(D)$
has the following representation (see, e.g., \cite[p.\ 90, Corollary
4.2.3]{Gray90},\cite{Rose94},\cite{Merhav08}):
\begin{equation}
R(D)=-\min_{\beta\ge 0}\min_Q\left\{\beta
D+\sum_{x\in\calX}p(x)\ln\left[\sum_{y\in\calY}q(y)e^{-\beta\rho(x,y)}\right]\right\}
\end{equation}
which, due to convexity in $\beta$ and concavity in $Q$, is equaivalent to
\begin{eqnarray}
R(D)&=&-\min_Q\min_{\beta\ge 0}\left\{\beta
D+\sum_{x\in\calX}p(x)\ln\left[\sum_{y\in\calY}q(y)e^{-\beta\rho(x,y)}\right]\right\}\nonumber\\
&=&-\min_{\beta\ge 0}\left\{\beta
D+\sum_{x\in\calX}p(x)\ln\left[\sum_{y\in\calY}q^*(y)e^{-\beta\rho(x,y)}\right]\right\},
\end{eqnarray}
and which, under the symmetry assumption, tells us that
for every point $(D,R)$ on the rate--distortion curve,
we have:
\begin{equation}
\label{base}
R=-
\min_{\beta\ge 0}\left\{\beta
D+\ln\left[\sum_{y\in\calY}q^*(y)e^{-\beta\rho(x,y)}\right]\right\}.
\end{equation}
Let $\beta^*$ achieve this minimum, i.e.,
\begin{equation}
-R=\beta^*D+\ln\left[\sum_{y\in\calY}q^*(y)e^{-\beta^*\rho(x,y)}\right],
\end{equation}
or, equivalently,
\begin{equation}
D(R)=-\frac{\ln\left[\sum_{y\in\calY}q^*(y)e^{-\beta^*\rho(x,y)}\right]+R}{\beta^*}
\end{equation}
Thus, clearly,
\begin{equation}
D(R)\le\max_{\beta\ge
0}\left\{-\frac{\ln\left[\sum_{y\in\calY}q^*(y)e^{-\beta\rho(x,y)}\right]+R}{\beta}\right\}=D_0(R),
\end{equation}
and so, it remains to show also the converse inequality, $D(R)\ge D_0(R)$. 
To this end, observe that eq.\ (\ref{base}) implies that
for every point $(D,R)$ on the rate--distortion function:
\begin{equation}
-R\le \beta D+
\ln\left[\sum_{y\in\calY}q^*(y)e^{-\beta\rho(x,y)}\right],
\end{equation}
holds for all $\beta\ge 0$ (with equality for $\beta=\beta^*$).
Equivalently, for all $\beta\ge 0$:
\begin{equation}
D\ge
-\frac{\ln\left[\sum_{y\in\calY}q^*(y)e^{-\beta\rho(x,y)}\right]+R}{\beta},
\end{equation}
and so,
\begin{equation}
D(R)\ge
\max_{\beta\ge
0}\left\{-\frac{\ln\left[\sum_{y\in\calY}q^*(y)e^{-\beta\rho(x,y)}\right]+R}{\beta}\right\}=D_0(R),
\end{equation}
thus proving that $D_0(R)=D(R)$.

\section{Conclusion}

In this short paper, we tried to convey the following messages:
(i) There is an intimate relationship between tree coding and the
statistical physics of the DPRM, which we believe, is interesting,
first of all, on
its own right. (ii) The statistical mechanical approach provides an
alternative way to prove the tree coding theorem. 
(iii) Existing results concerning the DPRM are harnessed
right away to provide almost--sure convergence to the distortion--rate
function of the source, thus strenghening the existing coding theorem,
at least under a certain symmetry condition. 

It is speculated that the various statistical mechanical techniques 
that were exercised in the DPRM model (cf.\ last paragraph
of Subsection \ref{phys}) and otherwise
may shed more light on ensemble performance analysis on this and other
information--theoretic settings of theoretical and practical interest.
This research direction is currently pursued further.

\end{document}